\newif\ifpreprint
\preprintfalse 

\preprinttrue 

\documentclass[12pt,a4paper]{article}
\usepackage[latin1]{inputenc}
\usepackage{osajnl}
\usepackage{amsmath,amssymb,amsfonts,array,tabularx,color,xspace,pifont,times,amscd,pstricks} 
\ifpreprint \relax \else \usepackage{endfloat}\fi

\newcommand{\mykeywords}{%
 Interferometry (120.3180);
Image reconstruction-restoration (100.3020); 
Inverse problems  (100.3190).}

\usepackage[bookmarksnumbered,bookmarksopen=true,pdfusetitle,%
             colorlinks=true,breaklinks,linkcolor=blue,citecolor=blue]{hyperref}%
\hypersetup{%
  pdfsubject={JOSAA 2008 paper by S. Meimon et al. on Interferometry imaging},%
  pdfkeywords={\mykeywords}}

\ifpreprint
\usepackage[firsttwo,bottomafter,dark,outline,timestamp,dvips]{draftcopy}
\draftcopyName{To be published in JOSA A}{60}

\fi

\newlength{\osacolsize}

\input{alphabet}   

\def\Ldushort{$L_2L_1$}

\def\Ldu{\Ldushort\XS}

\def\Lduw{$L_2L_1^w$\XS}

\newcommand{\yba}{\boldsymbol{y}_{\boldsymbol{\alpha}(t)}}

\newcommand{\SOO}{OLBI~}

\newcommand{\myij}[1]{\ensuremath{\left\{#1\right\}_{ij}}}

\newcommand{\citeb}[1]{[\onlinecite{#1}]}

\newcommand{\defeq}{\stackrel{\Delta}{=} }
\newcommand{\congr}{\stackrel{2\pi}{\equiv} }
\newcommand{\Cdag}{\ensuremath{\Clot^{\dag}}}

\newcommand{\data}[1]{\ensuremath{{#1}^{\RM{data}}}}
\newcommand{\bruit}[1]{\ensuremath{{#1}^{\RM{noise}}}}
\newcommand{\noise}[1]{\ensuremath{{#1}^{\RM{noise}}}}
\newcommand{\prior}[1]{\ensuremath{{#1}^{\RM{prior}}}}
\newcommand{\eker}[1]{\ensuremath{{#1}^{\RM{ker}}}}

\newcommand{\wrt}{w. r. t. }

\newcommand{\Nt}{\ensuremath{N_{\RM{t}}}}

\newcommand{\GB}[1]{\left\{#1\right\}}

\newcommand{\TF}[1]{\operatorname{FT}\left[#1\right]}

\newcommand{\IMA}{\operatorname{im}} 
\newcommand{\rk}{\operatorname{rank}}

\newcommand{\RE}[1]{\Re \operatorname{e} \GB{#1}}
\newcommand{\IM}[1]{\Im \operatorname{m} \GB{#1}}

\newcommand{\XXXX}[1]{} 

\newcommand{\begit}[1]{\begin{itemize}#1\end{itemize}}

\newcommand{\bs}[1]{\ensuremath{\boldsymbol{#1}}}

\newcommand{\bsr}{\boldsymbol{r}}
\newcommand{\bsR}{\boldsymbol{R}}

\newcommand{\bsB}{\boldsymbol{B}}
\newcommand{\bsBb}{\bar{\boldsymbol{B}}}

\newcommand{\bsx}{\boldsymbol{x}}

\newcommand{\bsO}{\boldsymbol{O}}

\newcommand{\bsH}{\boldsymbol{H}}

\newcommand{\bss}{\boldsymbol{s}}

\newcommand{\bsV}{\boldsymbol{V}}
\newcommand{\bsu}{\boldsymbol{u}}

\newcommand{\bsnu}{\boldsymbol{\nu}}
\newcommand{\bsmu}{\boldsymbol{\mu}}

\newcommand{\bsalpha}{\boldsymbol{\alpha}}
\newcommand{\bskappa}{\boldsymbol{\kappa}}

\newcommand{\bsphi}{\boldsymbol{\phi}}
\newcommand{\bsvarphi}{\boldsymbol{\varphi}}
\newcommand{\bsxi}{\boldsymbol{\xi}}

\newcommand{\bsa}{\boldsymbol{a}}

\newcommand{\bsM}{\boldsymbol{M}}
\newcommand{\bsy}{\boldsymbol{y}}

\newcommand{\Clot}{\boldsymbol{C}}
\newcommand{\bsCl}{\bs{\beta}}

\newcommand{\moy}[2]{\left\langle #2 \right\rangle_{#1}}

\newcommand{\Esp}[1]{E\left\{ #1\right\}}

\newcommand{\ie}{{\em i.e. }}
\newcommand{\eg}{{\em e.g. }}

\newcommand{\bbar}[1]{\bar{\bar{#1}}}

\newcommand{\beq}[1]{\begin{equation}#1\end{equation}} 
\newcommand{\bea}[1]{\begin{align}#1\end{align}} 
\newcommand{\bead}[1]{\begin{aligned}#1\end{aligned}} 
\newcommand{\beas}[1]{\begin{align*}#1\end{align*}}

\newcommand{\diag}[1]{\ensuremath{\textrm{Diag}\left\{#1\right\}}}
\newcommand{\var}{\ensuremath{\textrm{Var}}}

\newcommand{\cov}[1]{\ensuremath{\bsR_{#1}}}
\newcommand{\Id}{\ensuremath{\textrm{{\bf I}d}}}

\newcommand{\gauss}[2]{\sim \mathcal{N}\left(#1,#2\right)}

\newcommand{\RM}[1]{\mathrm{#1}}

\newcommand{\EXP}{\mathrm{e}}           
\newcommand{\I}{\mathrm{i}}           
\newcommand{\Abs}[1]{\left\vert #1\right\vert}
\newcommand{\abs}[1]{\vert #1\vert}
\newcommand{\Avg}[1]{\left\langle #1\right\rangle}

\newcommand{\T}{^{\mathrm{T}}}        

\newcommand{\J}{\mathcal{J}}

\newcommand{\Falt}{\mathcal{F}_{\alphab(t)}}

\newcommand{\so}{\Rightarrow}

\newcommand{\brad}[1]{\ensuremath{{{#1}_{\RM{rad}}^{\RM{n}}}}}
\newcommand{\btan}[1]{\ensuremath{{{#1}_{\RM{tan}}^{\RM{n}}}}}

\newcommand{\bby}{\left[\begin{array}{c}\brad{\bsy}\\\btan{\bsy}\end{array}\right]}
\newcommand{\bbys}{\bruit{\bbar{\bsy}}}

\newcommand{\refp}[1]{\ref{#1} page \pageref{#1}}

\begin{document}
\setlength{\osacolsize}{8.4cm}

\title{A self-calibration approach for optical long baseline interferometry imaging}

\author{Serge Meimon,$^{1,*}$ Laurent M.
  Mugnier$^{1}$ and Guy Le Besnerais$^{2}$}
\address{$^{1}$Office National d'Études et de Recherches
        Aérospatiales\\ Département d'Optique Théorique et Appliquée\\ BP 72,
        F-92322 Châtillon cedex, France}

\address{$^{2}$Office National d'Études et de Recherches
        Aérospatiales\\ Département Traitement de l'Information et Modélisation \\ BP 72,
        F-92322 Châtillon cedex, France}
\address{$^*$Corresponding author: lastname@onera.fr}

\begin{abstract} 
 Current optical interferometers are affected by unknown turbulent phases on
  each telescope. In the field of radio-interferometry, the self-calibration
  technique is a powerful tool to process interferometric data with missing
  phase information. This paper intends to revisit the application of
  self-calibration to Optical Long Baseline Interferometry (\SOO). We cast
  rigorously the \SOO data processing problem into the self-calibration
  framework and demonstrate the efficiency of the method on real astronomical
  \SOO dataset.
\end{abstract}

\ocis{\mykeywords}

\bigskip
\section{Introduction\label{sect:intro}}


Optical Long Baseline Interferometry (\SOO) aims to combine light collected by
widely separated telescopes to access angular resolutions beyond the
diffration limit of each individual aperture. Long-baseline interferometers
measure a discrete set of spatial frequencies of the observed object, or
Fourier data. Due to instrumental complexity, current interferometers
recombine only a few telescopes, and even several nights of
observation lead to a very limited number of Fourier data; moreover, due to
the atmospheric turbulence, it is very difficult to get reliable phase
information from ground based interferometry~\citeb{Principles-chap13}. Hence
\SOO has to deal with severe under-determination and missing phase
information.

The classical answer to under-determination is to use a parametric approach,
\ie to search for an object entirely described by a small set of parameters
(for instance a circular object with a parametric attenuation profile). With a
``good model'', such an approach allows a reliable and precise estimation of
astrophysical parameters. A good model should limit as much as possible the
number of free parameters, while allowing a description of all the object's
features, because parametric inversion cannot reveal unguessed features. The
$\chi^2$ fit is often used as a model quality diagnosis, since an inadequate
model will often result in a poor fit to the data, thus revealing that a new
model (with more parameters or different parameters) is needed. However, it
does not reveal which new model must be adopted.

As progress in instrumental issues gives access to better frequency coverage,
\ie to potentially finer descriptions of the object, the choice of the model
becomes more difficult. An alternate and complementary approach is then
non-parametric reconstruction, which we will call "optical long baseline
interferometric imaging" (OLBII). Imaging means that the object is described
by a large set of parameters, such as coefficients of the object's
decomposition in some spatial functional basis, while under-determination is
tackled by regularization tools. Imaging is useful to understand the structure
of a complex object when prior information is limited.

From the beginning, OLBII has been
influenced by the remarkable techniques developed in
radio-interferometry with very large baselines
(VLBI)~\citeb{Thompson-b-86}. For instance, the "WIPE" OLBII technique of
A. Lannes et al.\citeb{Lannes-a-97} is inspired by the
well-known CLEAN method~\citeb{Hogbom-74}. As regards the missing phase
problem, the self-calibration technique proposed in
radio-interferometry by Cornwell and Wilkinson~\citeb{Cornwell-81} underlies
recent works in OLBII~\citeb{Lannes-a-98}.

This paper intends to revisit the application of self-calibration to
\SOO. Our contribution is three-fold:
\begin{enumerate}
\item we cast rigorously the \SOO data processing problem into the
  self-calibration framework, with consideration of the second-order
  statistics of the noise;
\item we propose WISARD (for Weak-phase Interferometric Sample Alternating
  Reconstruction Device), a self-calibration algorithm dedicated to OLBII, which
  uses the proposed data model within a Bayesian regularization approach;
\item we demonstrate the efficiency of WISARD on real astronomical \SOO dataset.
\end{enumerate}

The paper is organized as follows: Section~\ref{sect:OIobs} describes the
observation model of \SOO, briefly presents a Bayesian approach and discusses
the main problems that are encountered because of the incomplete \SOO data.
Section~\ref{sect:myope_model} is devoted to the derivation of a specific
myopic model, which achieves a good approximation of the data model and leads
to self-calibration techniques. One such technique, \textsc{Wisard},
is proposed in Section~\ref{sect:WISARD}. Results of \textsc{Wisard} on
simulated and real astronomical datasets are presented in
Section~\ref{sec:soo-resultats}. Our conclusions are given in
Section~\ref{sect:conclusion}. Most mathematical derivations are gathered in
the Appendices.

\bigskip
\section{Realistic observables in optical long baseline interferometry\label{sect:OIobs}}

\subsection{Ideal interferometric data}

Here we describe the ideal data, \ie without aberrations, noise or turbulence
effects, produced by a \Nt-telescope interferometer observing a monochromatic
source with wavelength $\lambda$. The brightness distribution of the source is
denoted $x(\bsxi)$, $\bsxi$ being angular coordinates on the sky. Individual
telescopes $T_k$ of the interferometer are located at three-space positions
$\overrightarrow{OT}_k$, and we denote $\bsr_k(t)$ the projection of
$\overrightarrow{OT}_k$ onto $\mathcal{P}$, the plane normal to the pointing
direction. Because of the Earth's rotation, the pointing direction
  changes during an observing night, so these projected vectors are time
  dependent.

Each pair $(T_k,T_l)$ of telescopes yields a fringe pattern with a 2D spatial
frequency $\bsnu_{kl}(t)\defeq\frac{\bsu_{kl}(t)}{\lambda}, $ where
$\bsu_{kl}(t)$ is the \emph{baseline} \beq{\label{eq:ur}\bsu_{kl}(t)\defeq
  \bsr_l(t)-\bsr_k(t),} \ie the projection of the vector $\overrightarrow{T_k
  T_l}$ onto $\mathcal{P}$.

Measuring the position and contrast of these
fringes yields a phase $\data{\phi}_{kl}(t)$ and an amplitude
$\data{a}_{kl}(t)$, which can be grouped together in a complex visibility
\begin{equation}
\data{y}_{kl}(t)\defeq\data{a}_{kl}(t)\EXP^{\I\data{\phi}_{kl}(t)}.
 \label{eq:yaphi}
\end{equation}
According to
the Van Cittert-Zernike theorem \citeb{Goodman-85}, complex visibilities are
\emph{ideally} linked to the normalized Fourier Transform (FT) of $x(\bsxi)$ at the 2D
spatial frequency $\bsnu_{kl}(t)$ through
\begin{equation}
\data{y}_{kl}(t)=\eta_{kl}(t)\frac{\TF{x(\bsxi)}(\bsnu_{kl}(t))}{\TF{x(\bsxi)}(\bs0)}.
\label{eq-obs}
\end{equation}

The \textit{instrumental visibility} $\eta_{kl}(t)$ accounts from the many
potential sources of visibility loss: residual perturbations of the wavefront
at each telescope, differential tilts between telescopes, differential
polarization effects, non-zero spectral width, etc. In practice, the
instrumental visibility is calibrated on a star reputed to be unresolved by
the interferometer before the object of interest is observed, and is
compensated for in the pre-processing of the raw data. Thanks to this
calibration step, we replace $\eta_{kl}(t)$ by $1$ in equation~(\ref{eq-obs}).

For the sake of clarity, we consider a \textit{complete} \Nt-telescope array
in what follows, \ie one in which all the possible two-telescope baselines can
be formed simultaneously, and a non-redundant interferometer configuration,
where each baseline provides a different spatial frequency. Extension to
incomplete and redundant settings is straightforward. Thus, at each time $t$,
there are \beq{\label{eq:nb} N_b={\Nt \choose 2}=\frac{\Nt(\Nt-1)}{2}}
complex observation equations such as~(\ref{eq-obs}).

Let us briefly introduce the discretized observation model. The sought
brightness distribution $x$ is represented by the coefficients $\bsx$ of its
projection onto some convenient spatial basis (box functions, sinc's, wavelets,
prolate spheroidal functions, etc...). The normalized discrete-continuous
Fourier matrix $\bsH(t)$ maps the chosen discrete spatial representation into
the real-valued instantaneous frequency coverage $\{\bsnu_{kl}(t))\}_{1\leq
  k<l\leq \Nt}$, and we further define
\begin{align}
\left\{ \begin{aligned}
{\bsa}(\bsx,t)&\defeq \abs{\bsH(t)\bsx}\\
{\bsphi}(\bsx,t)&\defeq \arg\left\{\bsH(t)\bsx\right\}.
\end{aligned}\right.
\label{eq:aphix}
\end{align}

\subsection{Effect of atmospheric turbulence on short-exposure measurements}

At optical wavelengths, atmospheric turbulence affects phase measurements
through path length fluctuations. The statistics of these fluctuations can be
described by a time scale parameter, the \textit{coherence time} $\tau_0$, typically
around 10 milliseconds, and by a space scale parameter, the \textit{Fried
  parameter} $r_0$
\citeb{Fried-65}. We assume that the diameter of the elementary apertures is
small relative to the Fried parameter, or that each telescope is corrected from the effects of turbulence by adaptive optics. The remaining turbulent effects on the interferometric measurements can be seen as a delay
line between the two telescopes $T_k$ and $T_l$, which affects
short-exposure phase measurements through an additive \textit{differential piston} $\varphi_l(t)-\varphi_k(t)$:
\begin{equation}
  \data{\phi}_{kl}(t)=\phi_{kl}(\bsx,t)+\varphi_l(t)-\varphi_k(t) + \mbox{noise} \; [2\pi]
\label{eq:obsphi}
\end{equation}
or, in a matrix formulation:
\begin{equation}
\data{\bsphi}(t)=\bsphi(\bsx,t)+\bsB\bsvarphi(t) + \mbox{noise}\; [2\pi]
\label{eq:Bphi}
\end{equation}
where $N_b\times \Nt$ operator $\bsB$, called the \emph{baseline
  operator}, is defined in Appendix~\ref{app:BandC}.

Because the differential pistons are zero-mean, one might think that the object
phase $\bsphi(\bsx,t)$ could be recovered from (\ref{eq:Bphi}) by averaging over many realizations
of the atmosphere. However, for a long baseline relative to the Fried
parameter, the optical path difference between apertures introduced by
turbulence may be very much greater than the observation wavelength and thus
lead to random pistons much larger than $2\pi$. The $2\pi$-wrapped
perturbation that affects the phase (\ref{eq:Bphi}) is then practically
uniformly distributed in $[0, 2\pi]$. In consequence, averaging the 
short-exposure phases measurements (\ref{eq:Bphi}) does not improve the signal-to-noise
ratio.

In \emph{phase referencing} techniques (see
\citeb{Principles-chap9}), the turbulent pistons are measured in order to subtract
them in (\ref{eq:Bphi}). However powerful and promising, these methods
require specific hardware and are not feasible for all sources. The only other way
to obtain exploitable long-exposure data then is to form piston-free short-exposure
observables \textit{before} the averaging.

\subsection{Piston-free short-exposure observables}
Piston-free short-exposure phase observables are quantities $f(\data{\bsphi}(t))$ in
which the turbulent term $\bsB\bsvarphi(t)$ cancels out:
\begin{equation}
f(\data{\bsphi}(t))=f(\bsphi(\bsx,t)+\bsB\bsvarphi(t))=f(\bsphi(\bsx,t)).
\label{eq:fphi}
\end{equation}
For an interferometric array of 3 telescopes or more, the \textit{closure
  phases} \citeb{Jennison-58} are one famous example, in which $f$ is a linear
operator performing triple-wise summation of the phases. For any set of three
telescopes $(T_k,T_l,T_m)$ the short-exposure visibility phase data are
\begin{align}
\left\{ \begin{aligned}
\data{\phi}_{kl}(t)&=\phi_{kl}(\bsx,t)+
\varphi_{l}(t)-\varphi_{k}(t) + \mbox{noise}\; [2\pi]\\
\data{\phi}_{lm}(t)&=\phi_{lm}(\bsx,t)+
\varphi_{m}(t)-\varphi_{l}(t) + \mbox{noise}\; [2\pi]\\
\data{\phi}_{mk}(t)&=\phi_{mk}(\bsx,t)+
\varphi_{k}(t)-\varphi_{m}(t) + \mbox{noise}\; [2\pi]
\end{aligned}\right.
\end{align}
and the turbulent pistons cancel out in the closure phase defined by :
\begin{eqnarray}
\label{eq:cloture}
\begin{aligned}
\data{\beta}_{klm}(t)&\defeq\data{\phi}_{kl}(t)+
\data{\phi}_{lm}(t)+\data{\phi}_{mk}(t) + \mbox{noise}\; [2\pi]\\
&=\phi_{kl}(\bsx,t)+\phi_{lm}(\bsx,t)+\phi_{mk}(\bsx,t) + \mbox{noise}\; [2\pi]\\
&\defeq\beta_{klm}(\bsx,t) + \mbox{noise}\; [2\pi]\;.
\end{aligned}
\end{eqnarray}

We have the following properties :
\begin{itemize}
\item the set of all three-telescope closure phases that can be formed using a
  complete array is generated by the ($\Nt-1)(\Nt-2)/2$ closure phases
  $\data{\beta}_{1kl}(t),\,k<l$, \ie the closure phase which include telescope
  $T_1$ (indeed,
    $\data{\beta}_{klm}=\data{\beta}_{1kl}+\data{\beta}_{1lm}-\data{\beta}_{1km}$).
  In what follows, these canonical closure phases are grouped together in a
  vector $\data{\betab}$ and $\Clot$ denotes the linear closure operator such
  that $\Clot\data{\bsphi}=\data{\betab}$ (see appendix \ref{app:BandC}).
\item if $f$ is a continuous differentiable function verifying
  property~(\ref{eq:fphi}), then
\begin{equation}
f(\bsphi)=g(\Clot\bsphi),
\label{eq:fdonneg}
\end{equation}
where $g$ is some continuous differentiable function. In other terms, there is
essentially \textit{ no operator other than the closure operator} that cancels
out the effect of turbulence on short-exposure visibility phases (this
property holds only in the monochromatic case).
\end{itemize}
The proof of the second property is given in appendix \ref{app:proof}. 

\subsection{Long-exposure observables data model\label{sect:data_model}}
To minimize the effect of noise, one is led to average short-exposure
measurements, into long-exposure observables, chosen so that they are
asymptotically unbiased. The averaging time must be short enough \wrt the
earth rotation so that the baseline does not change, and long enough to reach
an acceptable Signal-to-Noise Ratio (SNR). The averaged quantities are
generally:
\begin{itemize}
\item averaged \emph{squared amplitudes} $\data{\bss}(t)=\moy{\tau}{\data{\bsa}(t+\tau)^2}$, 
\item averaged \emph{bispectra}
  \mbox{$\data{\bsV}_{1kl}(t)=\moy{\tau}{\data{y}_{1k}(t+\tau)\cdot\data{y}_{kl}(t+\tau)\cdot\data{y}_{l1}(t+\tau)},\,k<l$}.
 \end{itemize}
 Squared amplitudes are preferred to amplitudes because their bias can be
 estimated and subtracted from the data. Short-exposure bispectra are
 continuous differentiable functions verifying property~(\ref{eq:fphi}), and
 so correspond to a particular choice of $g$ in (\ref{eq:fdonneg}). In absence
 of noise, the
 averaged bispectrum amplitudes are redundant with the averaged
 squared amplitudes. Although they should be useful in low SNR conditions,
 averaged bispectrum amplitudes are not considered in what follows. The
 averaged bispectrum phases $\data{\beta}_{1kl}(t),\,k<l$ constitute unbiased
 long-exposure closure phase estimators. As such, they are linked to the
 object phases $\bsphi(\bsx,t)$ through:
\begin{equation}
\data{\bsCl}(t)=\Clot\bsphi(\bsx,t) + \mbox{noise} \; [2\pi]
\label{eq:clot}
\end{equation}
It is shown in appendix~\ref{app:BandC} that the kernel of the closure
operator $\Clot$ is of dimension $(\Nt-1)$. Hence equation~(\ref{eq:clot})
implies that optical interferometry through turbulence has to deal with a
partial phase information. This result can also be obtained by counting up
phase unknowns for each instant of measurement $t$: there are $\Nt(\Nt -1)/2$
unknown object visibility phases and $(\Nt-1)(\Nt -2)/2$ observable
independent closure phases, which results in $(\Nt-1)$ missing phase data. As
well known in the radio-interferometric community, the more apertures in the
array, the smaller the proportion of missing phase information will be.

The long-exposure observables considered in this paper are noisy {squared
  amplitudes} $\data{\bss}(t)$ and closure phases $\data{\bsCl}(t)$. The
only statistics usually available are the variances for each
observable (as, for instance, in the OIFITS data exchange format~\citeb{OIFITS}).
The assumed noise distribution is consequently 0-mean
white Gaussian:
\bea{\label{eq:model}\left\{\begin{aligned}
\data{\bss}(t)&={\bsa^2}(\bsx,t)+\bruit{\bss}(t),\;\;\;\bruit{\bss}(t)\gauss{\zerob}{\bsR_{\bss(t)}}\\
\data{\bsCl}(t)&={\Clot\bsphi}(\bsx,t)+\bruit{\bsCl}(t) \; [2\pi],\;\;\;\bruit{\bsCl}(t)\gauss{\zerob}{\bsR_{\bsCl(t)}}
\end{aligned}\right.
} The matrices $\Rb_{\sb(t)}$ and $\Rb_{\betab(t)}$ are diagonal, with
variances related to the integration time, although correlations may be
produced by the use of the same reference stars in the calibration process
\cite{Perrin-a-03} .

\subsection{Bayesian reconstruction methods\label{sect:Bayes}}

This approach first forms the anti-log-likelihood according to the model (\ref{eq:model})
\begin{equation}
\label{jmira}\data{J}(\bsx)=\sum\limits_{t}\data{J}(\bsx,t)=\sum\limits_{t}\chi^2_{\sb(t)}(\bsx)+\chi^2_{\betab(t)}(\bsx)
\end{equation}
where $\chi^2_{\sb(t)}(\bsx)$ denotes the $\chi^2$ statistic $
\left(\data{\sb}(t)-
  {\ab}^2(\bsx,t)\right)\T\Rb_{\sb(t)}^{-1}\left(\data{\sb}(t)-{\ab}^2(\bsx,t)\right)
$. Closure terms $\chi^2_{\betab(t)}(\bsx)$ are a weighted quadratic
distance between complex phasors \cite{Haniff91} instead of a Chi-2 statistic over closure
phase residuals. One then associates $\data{J}$ with a regularization term to
account for the incompleteness of the data in such inverse problems and
minimizes the composite criterion
\begin{equation}
\label{eq:crit-soo-regul}
J(\bsx) = \data{J}(\bsx) + \prior{J}(\bsx)
\end{equation}
under the following constraints:
\begin{align}
\begin{aligned}
\forall (p,q),\; x(p,q)&\geq 0\\
\sum\limits_{p,q} x(p,q)&=1.
\end{aligned}
\label{eq:contraintes}
\end{align}
The first requires positivity of the sought object, the second is a constraint
of unit flux. Indeed, fringe visibilities are by definition flux-normalized
quantities (\ie normalized by the Fourier transform of the object at the null
frequency, see Eq.~\ref{eq-obs}), so the data are independent of the total
flux of the sought object (of course an interferometer is sensitive to
  the total flux of the source, but this last value is not contained in the
  fringe visibility itself).

The regularization term $\prior{J}$ is chosen to enforce some properties of
the object which are known a priori (smoothness, spiky
behavior, positivity, etc.) and should also ease the minimization. Simple and
popular regularization terms are convex separable penalizations of the object
pixels (\ie \textit{white priors}) or of the object spatial derivatives (for
instance first-order derivative or gradient). In what follows, we quickly
describe the prior terms used in this paper. These priors are more extensively
described and compared in ~\citeb{Besnerais-a-08}. For a general review on
regularization, see ~\citeb{Demoment-89}.
  
Entropic priors belong to the family of white priors and often allow to obtain
a clean image while preserving its sharp spiky features, whereas quadratic
penalization tends to soften the reconstructed map. The white quadratic-linear
(or \Lduw) penalization given by:
\beq{\label{l1l2w}L_2L_1^w(\bsx)=\delta^2\sum\limits_{p,q}\frac{\bsx(p,q)}{s\delta}-\ln\left(1+\frac{\bsx(p,q)}{s\delta}\right)}
that we use in section~\ref{sec:soo-resultats} leads to a kind of entropic
regularization, in the sense of~\citeb{Nityananda-82}. We propose a nominal
setting of the two parameters $\delta$ and $s$:
\beq{\label{auto}s=1/N_{\RM{pix}};\;\;\;\delta=1.}

As regards regularization based on the object's spatial derivatives, we shall
consider here only quadratic penalization, but convex quadratic-linear \Ldu
penalization functions could also be invoked.

Ref.~\citeb{Thiebaut-a-03} is one of the works that adopt such a Bayesian approach for
processing optical long baseline interferometry, using a constrained local descent method to
minimize (\ref{eq:crit-soo-regul}). A convex data criterion $J$, i.e.
  such that $J(k\cdot x_1+(1-k)\cdot x_2) \le k\cdot J( x_1)+(1-k)\cdot
  J(x_2),\;\forall x_1,x_2,\;\forall k \in [0,1]$,  has no local minima, which makes the
  minimization much easier. Unfortunately, the criterion $J$ is non-convex. To be more precise, the
difficulty of the problem can be summed up as follows:
\begin{enumerate}
\item[(i)] The small number of Fourier coefficients makes the problem
  under-determined. Here the regularization term and the positivity constraint
  can help by limiting the high frequencies of the reconstructed
  object~\citeb{Lannes-a-98}.
\item[(ii)] Closure phase measurements implies missing phase
  information and makes the Fourier synthesis problem non-convex. Adding a
  regularization term does not generally correct the
  problem~\citeb{Navaza-a-92}.
\item[(iii)] Phase and modulus measurements with additive Gaussian noise leads
  to a non-Gaussian likelihood and a non-convex log-likelihood w.r.t. $\bsx$.
  As a consequence, even with no missing phases, some approximation of the
  real observable statistics is necessary to get a convex data fidelity term.
  This data conversion from polar to Cartesian coordinates, which is commonly
  used in the field of radar processing~\citeb{Bar-Shalom}, has been studied
  only recently in \SOO~\citeb{Meimon-a-05}: see section~\ref{sect:approx}.
\end{enumerate}

These characteristics imply that optimizing $J$ by a local descent algorithm
can only work if the initialization selects the "right" valley of the
criterion. The design of a good initial position is very case-dependent, and
will not be extensively addressed here. The other key aspects are then the followed path,
\ie the minimization method, and the shape of the function to minimize,
\ie the behavior of the criterion $\bsx \mapsto J(\bsx)$. This paper addresses both aspects:
\begin{itemize}
\item we design a specific \SOO criterion $\J(\bsx,\bsalpha)$ where two sets of
  variables appear explicitly, one in the spatial domain $\bsx$, describing
  the sought object, and another in the Fourier phase domain $\bsalpha$, which
  accounts for the missing phase information. This specific criterion is
  designed to solve (iii), \ie so that for a known $\bsalpha$, the criterion
  is convex \wrt $\bsx$. In other words, if we had all the complex visibility
  phase measurements instead of just the closure phases, our criterion $\bsx \mapsto \J(\bsx,\bsalpha)$
  \textit{would be convex};
\item we adopt an alternate minimization method, working on the two sets of
  variables.
\end{itemize}
This approach can be related to "myopic" approaches of some inverse problems,
where missing data concerning the instrumental response are modeled and sought
for during the inversion~\citeb{Mugnier-l-08a}. Alternate minimization methods
are inspired by self-calibration methods in radio-interferometry, and have
been used in optical interferometry by Lannes et al.~\citeb{Lannes-a-98}.
However, the criterion used in Ref.~\citeb{Lannes-a-98} was essentially
imported from radio-interferometry and does not match the \SOO data
model~(\ref{eq:model}). Our main contribution is to derive a criterion which
accounts for the data model (\ref{eq:model}), while allowing an efficient
alternate minimization. This construction is the subject of the next section.

\section{An equivalent myopic  model for self-calibration\label{sect:myope_model}}

The aim of this section is to approximate the data model of equation~(\ref{eq:model}):
\begin{eqnarray}
\data{\bss}(t)&=&{\bsa^2}(\bsx,t)+\bruit{\bss}(t),\;\;\;
\bruit{\bss}(t)\gauss{\zerob}{\bsR_{\bss(t)}}\label{eq:modsquare}\\
\data{\bsCl}(t)&=&{\Clot\bsphi}(\bsx,t)+\bruit{\bsCl}(t) \; [2\pi],\;\;\;\bruit{\bsCl}(t)\gauss{\zerob}{\bsR_{\bsCl(t)}}\label{eq:modclot}
\end{eqnarray}
by a myopic linear model with additive complex Gaussian noise of the following form:
\bea{\label{mody}\data{\bsy}(t)=\Falt\cdot {\bsH(t)}\bsx+\bruit{\bsy}(t)}
where operator $\cdot$ denotes componentwise multiplication, and $\Falt$ is a vector of phasors depending on phase aberration parameters $\alphab(t)$, which are defined in Sec.~\ref{sect:phase}.
This will be done in three steps:
\begit{
\item Sec.~\ref{sect:amp} is devoted to the derivation of the observation model for
the pseudo amplitude term $\data{\bsa}(t)$ from~(\ref{eq:modsquare}); 
\item Sec.~\ref{sect:phase} is devoted to the derivation of the observation model for
the pseudo phase term $\data{\bsphi}(t)$ from~(\ref{eq:modclot}); 
\item Sec.~\ref{sect:approx} shows how to combine pseudo phase and pseudo
  amplitude models in a complex model such as equation (\ref{mody}) while solving
  problem (iii) of Sec.~\ref{sect:Bayes}.
}

\subsection{Pseudo amplitude data model\label{sect:amp}}

In Eq.~(\ref{eq:modsquare}), we have supposed a Gaussian distribution for $\data{\bss}(t)$ around
${\bss}(\bsx,t)$, which is questionable, since squared amplitudes should be
non-negative. However, such a statistic model is acceptable provided that the
probability of a negative component of $\data{\bss}(t)$ is very weak. For
uncorrelated measurements, this assumption correspond to mean values much
greater than the corresponding standard deviation.
Appendix~\refp{app:racgauss} shows how to build the mean and covariance matrix
of the square root of such a distribution. The mean vector is taken as the
pseudo amplitude data $\data{\bsa}(t)$, and the covariance matrix called
$\bsR_{\bsa(t)}$.

The observation model (\ref{eq:modsquare}) can then be approximated by
the following amplitude pseudo data model:
\bea{\label{modamp2}\data{\bsa}(t)={\bsa}(\bsx,t)+\bruit{\bsa}(t),\;\;\;\bruit{\bsa}(t)\gauss{\zerob}{\bsR_{\bsa(t)}}.}

\subsection{Pseudo phase data model\label{sect:phase}}

We start from a generalized inverse solution to the phase
closure equation of (\ref{eq:modclot}). The generalized inverse $\Clot^\dag$ of
$\Clot$, defined by $\Clot^\dag\defeq \Clot\T\left[\Clot\Clot\T\right]^{-1}$, is such that $\Clot\Clot^\dag=\Id$. By applying it on all the terms of~(\ref{eq:modclot}), we obtain
\begin{equation}
\Cdag\data{\bsCl}(t)={\Cdag\Clot\bsphi}(\bsx,t)+\Cdag\bruit{\bsCl}(t)+2\pi\,\Cdag\kappab
\end{equation}
where $\kappab$ is a vector of integers to account for the fact that each phase component is measured modulo $2\pi$.
We define
\begin{eqnarray}
\data{\bsphi}(t)&\defeq&\Clot^\dag\data{\bsCl}(t)\label{recphi1}\\
\eker{\bsphi}(t)&\defeq&(\Cdag\Clot - \Id)\bsphi(\bsx,t)+2\pi\,\Cdag\kappab
\label{recphi2}
\end{eqnarray}
and obtain
\bea{\data{\bsphi}(t)= \bsphi(\bsx,t)+ \eker{\bsphi}(t)+\Clot^{\dag}\bruit{\bsCl}(t)}

Vector $\eker{\bsphi}(t)$ belongs to the $2\pi$-wrapped kernel of operator $\Clot$ :
  \beas{\Clot\eker{\bsphi}(t)
    &=(\underbrace{\Clot\Cdag}_{=\Id}\Clot-\Clot)\bsphi(\bsx,t)+
    2\pi\underbrace{\Clot\Cdag}_{=\Id}\kappab\\
    &=2\kappab\pi \\
    &= \zerob \; [2\pi] }

  As shown in appendix~\ref{app:intamb}, if $\eker{\bsphi} = 0\; [2\pi]$, there
  exists a real vector $\bsalpha(t)$ of dimension $\Nt-1$ such that
  $\eker{\bsphi}(t) = \bsBb\bsalpha(t)\; [2\pi]$, where $\bsBb$ is obtained by
  removing the first column of operator $\bsB$. So we have:
  \bea{\label{kerclot}\data{\bsphi}(t) = \bsphi(\bsx,t)+
    \bsBb\bsalpha(t)+\Clot^{\dag}\bruit{\bsCl}(t)\; [2\pi]}
  Now the problem is that $\Clot^{\dag}\bruit{\bsCl}(t)$ is a zero mean random
  vector with a \textit{singular covariance matrix}
$$\bsR_{\bsphi(t)}^0\defeq\Clot^\dag\bsR_{\bsCl(t)}\Clot^{\dag \RM{T}}.$$
To obtain a strictly convex log-likelihood, we have to approximate this term
by a proper Gaussian vector $\bruit{\bsphi}(t)$, with an invertible covariance
matrix $\bsR_{\bsphi(t)}$ chosen so as to correctly fit the second order
statistics of the noise in the phase closure measurement equation
(\ref{eq:modclot}). This last requirement can be written as the following
equation:
\begin{equation}
    \Clot\bsR_{\bsphi(t)}\Clot\T = \bsR_{\bsCl(t)}.
\label{eq:conditionCov}
\end{equation}
In other words, we are led to choose an invertible covariance matrix $\bsR_{\bsphi(t)}$ so as to mimic the statistical behavior of the closures, which is expressed by (\ref{eq:conditionCov}).

We propose to modify matrix $\bsR_{\bsphi(t)}^0$ by setting its non diagonal components to 0, \ie
to use the following diagonal matrix:
\bea{\myij{\bsR_{\bsphi(t)}}=\,\left\{\bead{&3\cdot\myij{\bsR_{\bsphi(t)}^0}&\textrm{
        if }&
      \;i=j\\
      &0 &\textrm{ if }&\;i\ne j}\right. .\label{recphi_cov}} The factor 3
allows us to preserve the total weight of the phase term in the log-likelihood
by satisfying the condition:
$$\sum\limits_{i,j}\Abs{\myij{\bsR_{\bsphi(t)}}}=\sum\limits_{i,j}\Abs{\myij{\bsR_{\bsphi(t)}^0}}.$$
There are several ways of choosing $\bsR_{\bsphi(t)}$, and we propose
this particular choice without claiming it is optimal. Note that the myopic model
derived in what follows can accomodate to any choice of a proper (\ie invertible)
covariance matrix $\bsR_{\bsphi(t)}$.

With equations (\ref{recphi1}), (\ref{kerclot}) and (\ref{recphi_cov}), we obtain the visibility phase pseudo data model: 
\bea{\label{modphi2}\data{\bsphi}(t) =
  {\bsphi}(\bsx,t)+\bsBb\bsalpha(t)
  +\bruit{\bsphi}(t) \; [2\pi],\;\;\;\bruit{\bsphi}(t)\gauss{\bs0}{\bsR_{\bsphi(t)}}.}

\subsection{Pseudo complex visibility data
  model\label{sect:approx}}
Gathering equations (\ref{modamp2}) and (\ref{modphi2}), we have finally approximated the
data model (\ref{eq:modsquare}-\ref{eq:modclot}) by
\bea{\label{modpol}\left\{\bead{
      \data{\bsa}(t)&={\bsa}(\bsx,t)+\bruit{\bsa}(t),\;\;\;&\bruit{\bsa}(t)\gauss{\zerob}{\bsR_{\bsa(t)}}.\\
      \data{\bsphi}(t)&=
      {\bsphi}(\bsx,t)+\bsBb\bsalpha(t)+\bruit{\bsphi}(t)\; [2\pi],\;\;\;&\bruit{\bsphi}(t)\gauss{\bs0}{\bsR_{\bsphi(t)}}.}\right.}
We form pseudo-complex visibility measurements $\data{\bsy}(t)$ defined by:
\bea{\data{\bsy}(t)\defeq\data{\bsa}(t) \cdot \EXP^{\I\data{\bsphi}(t)}.} 
The approach proposed in~\citeb{Meimon-a-05}, which we recall and
generalize in Appendix~\ref{app:KL},
is based on an approximated complex visibility data model 
\bea{\label{mody2}\data{\bsy}(t)={\bsH(t)}\bsx\cdot\EXP^{\I\bsBb\bsalpha(t)}+\bruit{\bsy}(t)}
This is exactly the sought model stated the beginning of the present section in
equation (\ref{mody}), with $\Falt=\EXP^{\I\bsBb\bsalpha(t)}$. We now define the myopic observation model as follows: \begin{equation}
\bsy_m(\bsx,\bsalpha(t)) \defeq {\bsH(t)}\bsx\cdot\EXP^{\I\bsBb\bsalpha(t)}.
\end{equation}

\begin{figure}[h!]
\begin{center}
\includegraphics[width=.5\linewidth]{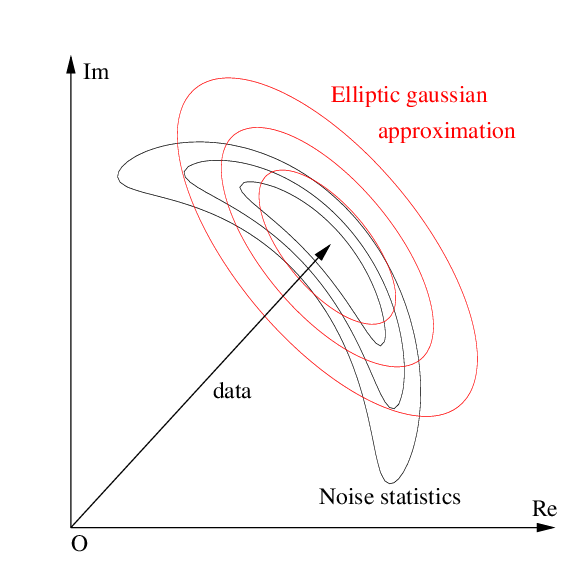}
\caption{Contour plots of a polar Gaussian distribution and of its Cartesian Gaussian approximation\label{fig:bananneandell}}
\end{center}
\end{figure}

As shown in Appendix~\ref{app:KL}, the mean value $\bruit{\bar{\bsy}}(t)$ and
covariance matrix $\bsR_{\bruit{\bsy}(t)}$ of the additive complex noise term
$\bruit{\bsy}(t)$ are carefully designed so that the corresponding data
likelihood criterion is convex quadratic \wrt the complex
$\bsy_m(\bsx,\bsalpha(t))$ while remaining close to the real non convex model.
To illustrate these properties, we consider one complex visibility and plot in
the complex plane the distribution of $\data{\bsy}(t)$ around
$\bsy_m(\bsx,\bsalpha(t))$ for the true noise distribution --- i.e. a polar
Gaussian noise in phase and modulus --- and our cartesian Gaussian
approximation (see Fig.~\ref{fig:bananneandell})
In particular, the ``elliptic'' covariance matrix we propose (which yields elliptic contour plots
in Fig.~\ref{fig:bananneandell}), is preferable to the more classical
``circular'' approximation that appears in previous contributions on \SOO~\citeb{Lannes-a-01}.
The latter can be described by half as many parameters as needed for the
elliptic one (one radius for a circle, instead of a short axis and a long axis for an
ellipsis), but is clearly less accurate~\citeb{Meimon-a-05} (such a noise
statistics description has also been investigated for the complex bispectra in the
OIFITS data exchange format~\citeb{OIFITS}).

From Eq.~(\ref{mody2}), we build a Chi-2 statistics over real and imaginary parts of the observation equation
\begin{eqnarray*}
  \lefteqn{
    \chi^2_{\bsy(t)}(\bsx,\bsalpha(t))\defeq
    \left[\begin{array}{c}\RE{\data{\bsy}(t)-\bsy_m(\bsx,\bsalpha(t))-\bruit{\bar{\bsy}}(t)}\\\IM{\data{\bsy}(t)-\bsy_m(\bsx,\bsalpha(t))-\bruit{\bar{\bsy}}(t)}\end{array}\right]\T
    \times} \\
  & {\bsR_{\bruit{\bsy}(t)}}^{-1} &
  \left[\begin{array}{c}\RE{\data{\bsy}(t)-\bsy_m(\bsx,\bsalpha(t))-\bruit{\bar{\bsy}}(t)}\\\IM{\data{\bsy}(t)-\bsy_m(\bsx,\bsalpha(t))-\bruit{\bar{\bsy}}(t)}\end{array}\right].
\end{eqnarray*}
And we finally propose the myopic goodness-of-fit criterion:
\beq{\label{jwiz}\data{\J}(\bsx,\bsalpha)=
  \sum\limits_{t}\data{\J}(\bsx,\bsalpha(t),t)=\sum\limits_{t}\chi^2_{\bsy(t)}(\bsx,\bsalpha(t))}
We can now design a myopic Bayesian approach to the reconstruction problem, by combining the
data term with a regularization term along the lines of Section~\ref{sect:Bayes}:
\begin{equation}
\label{eq:crit-myopic-regul}
\J(\bsx,\bsalpha) = \data{\J}(\bsx,\bsalpha) + \prior{J}(\bsx).
\end{equation}
The next section describes an alternate minimization technique applied to the regularized criterion~(\ref{eq:crit-myopic-regul}). 

 

\section{\textsc{Wisard}\label{sect:WISARD}}
In this section, we describe \textsc{Wisard}, standing for Weak-phase
Interferometric Sample Alternating Reconstruction Device, a self-calibration
method for OLBII.

\subsection{Global structure of \textsc{Wisard}}
\textsc{Wisard} is made of four major blocks:
\begit{
\item a first block recasts the raw data (i.e. closure phases and squared
  visibilities) in myopic data (i.e. phases and moduli) as described in
  sections~\ref{sect:amp} and~\ref{sect:phase};
\item a second "convexification block" computes a Gaussian approximation of the pseudo
  visibility data model as described in
  section~\ref{sect:approx};
\item a third block builds a guess for the object $\bsx$ and aberrations
  $\alphab$ (i.e. a good starting
  point);
\item finally, the self-calibration block performs the minimization of the regularized
  criterion~(\ref{eq:crit-myopic-regul}), under the constraints~(\ref{eq:contraintes}). It alternates
  optimization of the object for given aberrations, and optimization of the
  aberrations for the current object. }

\begin{figure}[h!]
\begin{center}
\includegraphics[width=.5\linewidth]{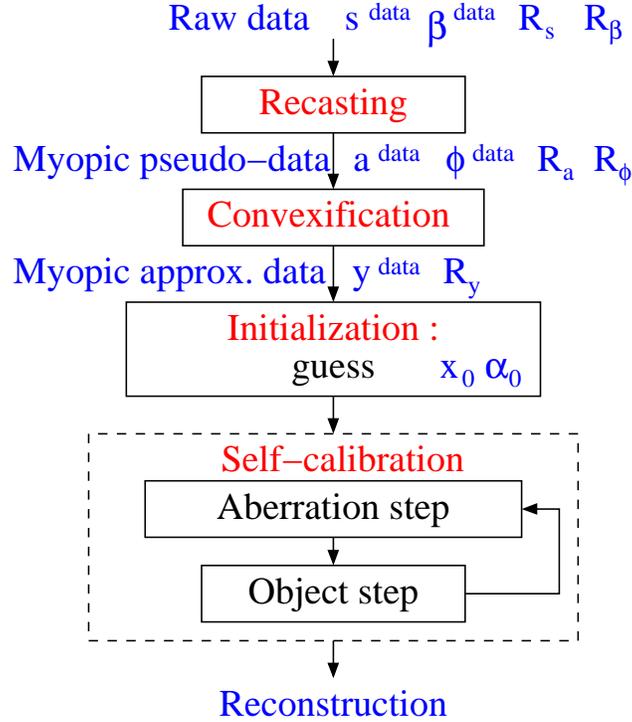}
\caption{\textsc{Wisard} algorithm loop\label{fig:boucle}}
\end{center}
\end{figure}

The structure of \textsc{Wisard} is sketched in Fig.~\ref{fig:boucle}. The
principles which underline the three first blocks of \textsc{Wisard} have been
described in previous Sections, while details on the self-calibration
minimization are gathered in the next one.

\subsection{Self-calibration block}
\paragraph{Minimization \wrt $\bsx$}
The criterion $\data{\J}(\bsx,\bsalpha)$ we have derived is quadratic hence convex w.r.t. the
object $\bsx$. Hence, the minimization versus $\bsx$ does not raise special
difficulties.

\paragraph{Minimization \wrt $\bsalpha$}
$\data{\J}(\bsx,\bsalpha)$ is the sum
of terms involving only measurements obtained at one time instant $t$ (equation~\ref{jwiz}):
$$\data{\J}(\bsx,\bsalpha)=\sum\limits_{t}\data{\J}(\bsx,\bsalpha(t),t)$$
Because the time between two measurements is much greater than the turbulence
coherent time (around 10 ms), aberrations $\bsalpha(t)$ at two different
instants are statistically independent. We can then solve separately for each
set of $\bsalpha(t)$, which dramatically reduces the complexity of the
minimization. The number of $\bsalpha(t)$ components to solve for is $(\Nt-1)$
and the minimization is delicate, as the criterion exhibits periodic
structures which have been studied in~\citeb{Lannes-a-01}.

However, exact minimization is affordable for a 3-telescope interferometric
array. In this case we have to perform several 2-parameter minimizations, and
each one can be efficiently initialized by an exhaustive search on a 2-D grid,
which ensures we avoid local minima. On the other hand, when $\Nt$ gets high
enough, \eg 6, then number of $\bsalpha(t)$ to solve for, \eg 5, gets small
compared to the number of closure phases available, \eg 15. With a 3-telescope
array, 2/3 of the phase information is missing, whereas with a 6-telescope
array, only 1/3 of the phase information is missing. In this last case, which
corresponds to the processing of synthetic data presented
Sec.~\ref{sec-traitement-donnees-synt}, the reconstructions were
straightforward, and no effects of the local minima in $\bsalpha$ were
witnessed.

In other words, coping with the ambiguities in $\bsalpha$, for instance with
the specific criterion proposed in~\citeb{Lannes-a-01}, may be
necessary only for $\Nt=4$ or $\Nt=5$. For $\Nt=3$, an exhaustive search is
possible, and for $\Nt \ge 6$, ambiguities in $\bsalpha$ do not have,
according to our experience, a major impact on reconstruction.

\paragraph{Starting point : object and aberration guess $\bsx_0$ and $\alphab_0$}
If a parametric model of the observed stellar source is not available, the
object starting point is a mean square solution, from which we extract the
positive part. The first step in the self-calibration block is a minimization
\wrt $\bsalpha$ for $\bsx=\bsx_0$. 

\section{Results\label{sec:soo-resultats}}

This section presents some results of processing by the 
\textsc{Wisard}~algorithm, with both synthetic and experimental data.

\subsection{Processing of synthetic data}\label{sec-traitement-donnees-synt}
\begin{figure}
\begin{center}\leavevmode
\includegraphics[width=.8\linewidth]{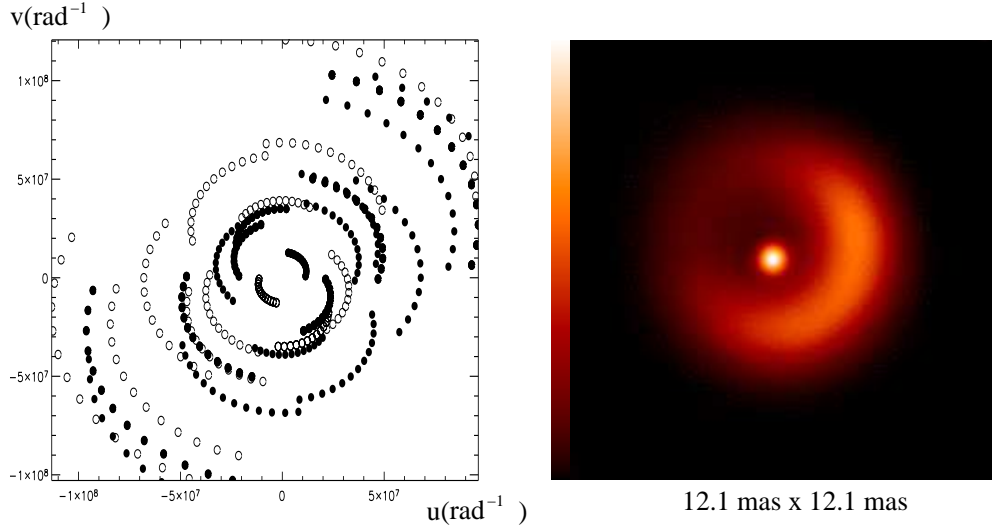}
\caption{Synthetic object (right) and frequency coverage (left) from the
Imaging Beauty Contest 2004
\label{fig:IBC_data}
}
\end{center}
\end{figure}

The first example takes synthetic interferometric data that were used in the
international Imaging Beauty Contest organized by P. Lawson for the
IAU~\citeb{BIC}. These data simulate the observation of the synthetic object
shown in figure~\ref{fig:IBC_data} with the NPOI~\citeb{NPOI-03} 6-telescope
interferometer. The corresponding frequency coverage, shown in
figure~\ref{fig:IBC_data}, has a structure in arcs of circles typical of the
\textit{super-synthesis} technique, which consists in repeating the
measurements over several nights of observation so that the same baselines
access different measurement spatial frequencies because of the Earth's
rotation. In total, there are 195 square visibility modules and 130 closure
phases, together with the associated variances.

\begin{figure}
\begin{center}\leavevmode
\includegraphics[width=\linewidth,clip=]{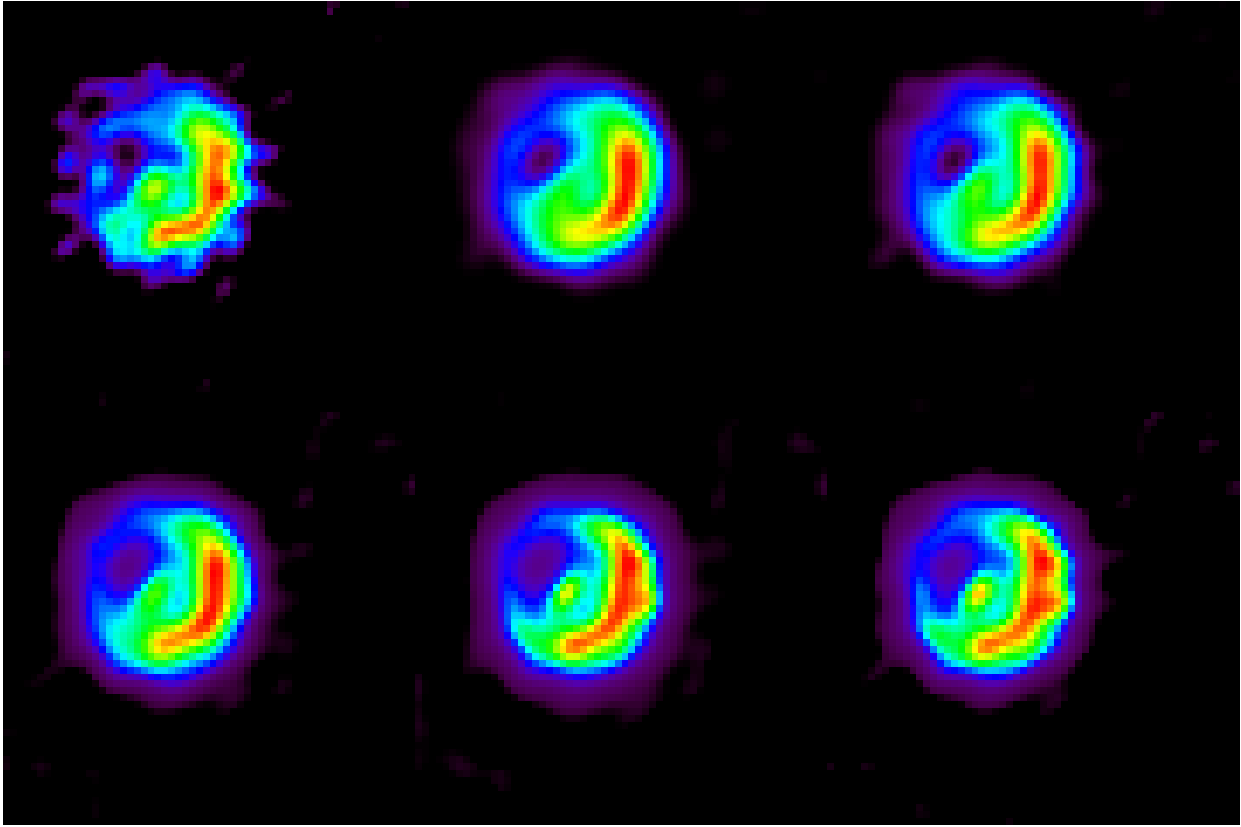}\,%
\caption{Reconstructions with \textsc{Wisard}. Upper row : under-regularized quadratic
model (left), over-regularized quadratic
model (center), quadratic model with correct regularization parameter (right).
Lower row : white \Lduw model with automatically set scale and delta parameters
(left), white \Lduw model with half scale (center), white \Lduw model with
half delta (right). Each image field is $12.1 \times 12.1$ mas.
}\label{fig:IBC_res_image}
\end{center}
\end{figure}

Six reconstructions obtained with \textsc{Wisard} are shown in
figure~\ref{fig:IBC_res_image}. On the upper row is a reconstruction using a
quadratic regularization based on a power spectral density model in $1/|u|^3$, for a weak, a
strong and a correct 
regularization parameter. The latter gives a satisfactory level of smoothing but does not
restore the peak in the center of the object. The peak is visible in the
under-regularized reconstruction on the left but at the cost of too high a
residual variance.

\begin{figure}
\begin{center}\leavevmode
  \includegraphics[width=\linewidth]{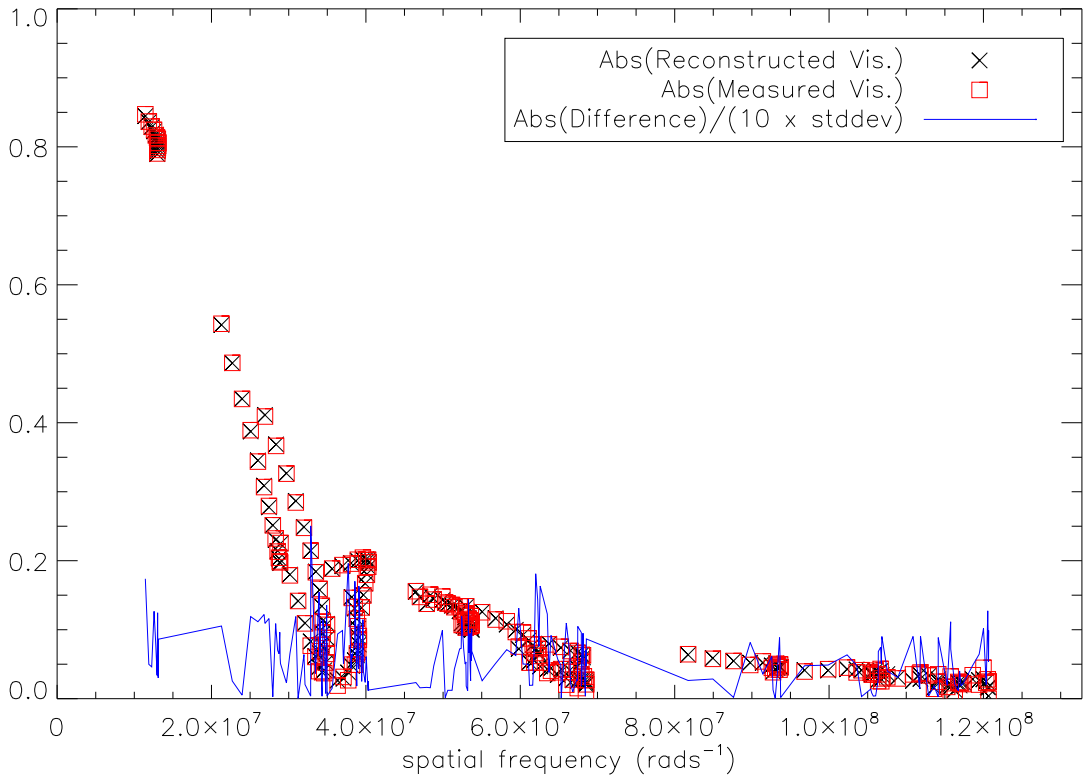}
  \caption{Goodness of fit at \textsc{Wisard} convergence.
  \label{fig:IBC-fit}
  }
  \end{center}
\end{figure}

The reconstruction presented on the lower row is a good trade-off between
smoothing and restoration of the central peak thanks to the use of the white
\Lduw prior term introduced in section~\ref{sect:Bayes}. The automatically set
parameters (eq.~\ref{auto}) are very satisfactory (left), and a light tuning (center and right)
allow an even better reconstruction. The goodness of fit of
the \Lduw reconstruction can be appreciated in figure~\ref{fig:IBC-fit}. The
red crosses show the reconstructed visibility moduli (i.e. of the FT of the
reconstructed object at the measurement frequencies) and the blue squares are the
moduli of the measured visibilities. The difference between the two, weighted
by 10 times the standard deviation of the moduli, is shown as the dotted line.
The mean value of this difference is $0.1, $ which shows a good fit (to within
1 $\sigma$).

\subsection{Processing of experimental data}\label{sec-traitement-donnees-exp}

Here, we present the reconstruction the star $\chi$ Cygni from experimental
data using the WISARD algorithm. The data were obtained by
S.~Lacour and S.~Meimon under the leadership of G.~Perrin during a measuring
campaign on the IOTA~interferometer \citeb{IOTA-06} in May 2005. As already
mentioned, each measurement has to be calibrated by observation of an object
that acts as a point source at the instrument's resolving power. The
calibrators chosen were HD~180450 and HD~176670.

$\chi$ Cygni is a Mira-type star, Mira itself being an example of such stars.
Perrin et al.~\citeb{Perrin-a-04} propose a model of Mira-type stars,
composed of a photosphere, an empty layer, and a thin molecular layer. The aim
of the mission was to obtain images of $\chi$ Cygni in the H band ($1{.}65$
microns $\pm 175nm$) and, in particular, to highlight possible assymmetric features
in the structure of the molecular layer.

\begin{figure}[htbp]
\begin{center}
\includegraphics[width=.85\linewidth]{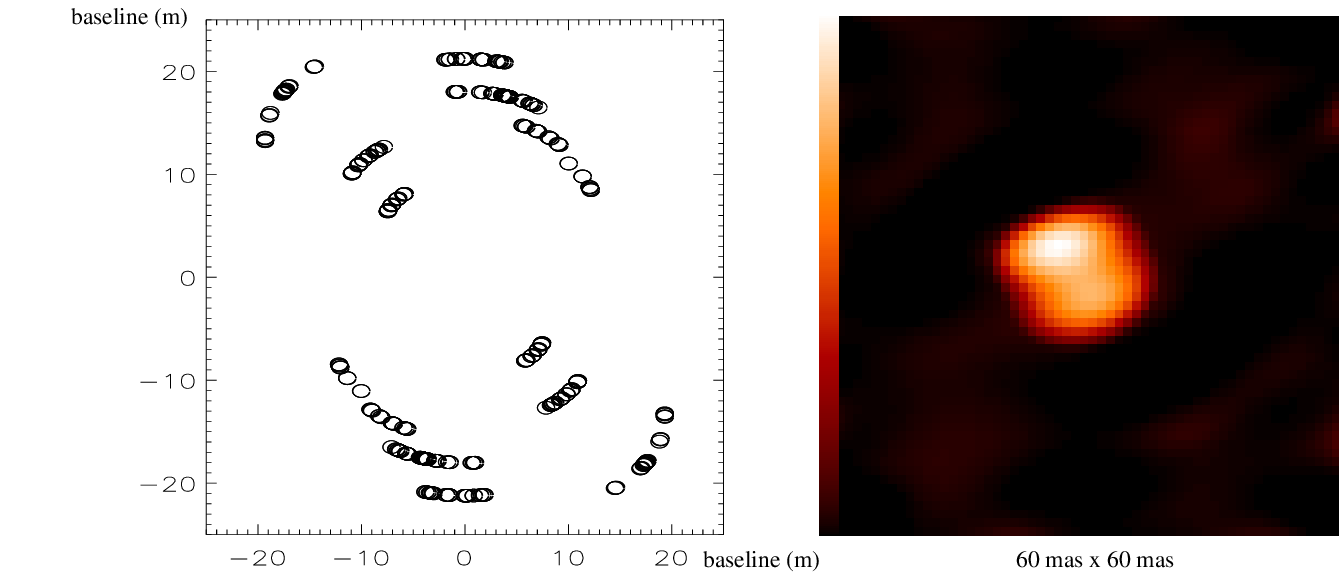}%
\caption{Frequency coverage (left) and reconstruction of the star
  $\chi$ Cygni (right).
\label{fig:recoIota}
}
\end{center}
\end{figure}

Figure \ref{fig:recoIota} shows, on the left, the $u-v$ coverage obtained,
i.e. the set of spatial frequencies measured, multiplied by the observation
wavelength. Because the sky is habitually represented with the west on the
right, the coordinates used are, in fact, $-u,v$. The domain of the accessible
$u-v$ plane is constrained by the geometry of the interferometer and the
position of the star in the sky. The "hour-glass" shape is characteristic of
the IOTA interferometer, and entails non-uniform resolution that affects the
image reconstruction, shown on the right. The reconstructed angular field has
sides of 60 milliarcseconds. In addition to the positivity constraint, the
regularization term used is the \Lduw term described in
section~\ref{sect:Bayes}.
The interested reader will find an astrophysical interpretation of this result
in \citeb{Lacour-t-07}.

\section{Concluding comments\label{sect:conclusion}}

We have proposed a complete and precise self-calibration approach to optical
interferometry image reconstruction. After pointing out the data model
specificities in the optical long baseline interferometry context, we have emphasized the
sources of under-determinations, which make a classical Bayesian criterion
descent method critical. Namely, the main problems are the phase under-determination
caused by turbulence effects, and, as noted only recently, the polar coordinate
structure of the data model. 

We have built a specially-designed approximate myopic data-model, in order to derive a self
calibration method. Special care was given to the design of the second order
statistics of the myopic model, an aspect which was ignored in previous related works. 

We have extended our previous work on polar data conversion~\cite{Meimon-a-05}
and proposed a convex approximation of the noise model which reduces the
number of local minima of the criterion to minimize.

We also addressed integer ambiguities induced by closure phase wrapping, which
are classical when dealing with phase data, and have discussed their impact on
the image reconstruction quality : for 3 telescope data, we have proposed an
exhaustive search method, and we have witnessed that these ambiguities do not
raise any particular problem when processing 6 or more telescope
interferometer data. Concerning the remaining 4-5 telescope case, the work by
Lannes \citeb{Lannes-a-01} should be worth investigating.On the other hand,
global minimization methods were left aside because of their intensive
computation needs. As computer performance increases, these
methods might be, in the years to come, an appropriate way to deal with local
minima.

All these developments allowed us to propose \textsc{Wisard}, a self-calibration method for
optical long baseline interferometry image reconstruction, and to demonstrate its efficiency
on simulated data. 

Finally \textsc{Wisard} was also used to successfully process real
astronomical \SOO datatsets. These
results were made possible thanks to a close partnership with the astronomers
Sylvestre Lacour and Guy Perrin of the Observatoire de Paris Meudon, whithin
the PHASE group (ONERA/LESIA). Indeed, an accurate astronomical model of the
observed stellar object is a precious guideline for reconstructing a complex
image from optical long baseline interferometric data. To the authors point of
view, such a collaboration is essential to the success of OLBII techniques.

\appendix

\section{The baseline and closure operators $\Clot$ and
  $\bsB$\label{app:BandC}}
Let $\Nt$ be the number of telescopes of the interferometric array. We have
the following definitions: \bea{\label{props}
  \bsB_2&\defeq\left[\begin{array}{cc}-1&1\end{array}\right]\\
  \bsB_{\Nt}&\defeq\left[\begin{array}{c|c}-\bs{1}_{\Nt-1}&\Id_{\Nt-1}\\\hline\bsO&\bsB_{\Nt-1}\end{array}\right]\\
\Clot_{\Nt}&\defeq\left[\begin{array}{c|c}-\bsB_{\Nt-1}&\Id_{\frac{(\Nt-1)(\Nt-2)}{2}}\end{array}\right]
}
for $\Nt\ge3$. 

In what follows, we prove that $ker\Clot=\IMA\bsB$.

We have $\Clot_{\Nt}\bsB_{\Nt}=\bs0$, so \bea{\label{equa}\IMA\bsB
  \subset\ker\Clot} It is straightforward to prove by recurrence that
$\bsB_{\Nt}\cdot\bs{1}_{\Nt}=0$, which yields $\rk \bsB_{\Nt} \le \Nt-1$.
Because $\bsB_{\Nt}$ contains $\Id_{\Nt-1}$ we gather: \bea{\label{equb}\dim\IMA\bsB\defeq\rk
  \bsB =\Nt-1.}

$\Clot_{\Nt}$ contains $\Id_{\frac{(\Nt-1)(\Nt-2)}{2}}$, which yields $\rk
\Clot_{\Nt} \ge \frac{(\Nt-1)(\Nt-2)}{2}$, or \bea{\label{equc}\dim\ker
\Clot_{\Nt} \le \Nt-1}
With (\ref{equa}), (\ref{equb}) and \ref{equc}), we gather: 
\beq{\label{lannes}\ker\Clot=\IMA\bsB }

\section{Characterization of the baseline phase independent
  operators\label{app:proof}}
Here, we prove that any continuous differentiable function $f$  verifying
property~\ref{eq:fphi}
$$f(\bsphi+\bsB\bsvarphi)=f(\bsphi),\;\forall(\bsphi, \bsvarphi) $$
is such that  $f(\bsphi)=g(\Clot\bsphi)$.
$\Clot$ has more columns
than rows, so its pseudo-inverse is defined by $\Cdag\defeq\Clot\T\left[\Clot\Clot\T\right]^{-1}$
and verifies 
 \beq{\label{eq:ccdag}\Clot \Cdag=\Id}
 and thus
\beas{\Clot \Cdag\Clot-\Clot=0 & \so \Clot \left(\Cdag\Clot\bsphi-\bsphi\right)=0,
    \;\forall \bsphi\\
 &\stackrel{\textrm{(\ref{lannes})}}{\so} \exists \bsvarphi, \left(\Cdag\Clot\bsphi-\bsphi\right)=\bsB\bsvarphi,
     \;\forall \bsphi\\
 &\so \exists \bsvarphi,\bsphi= \Cdag\Clot\bsphi-\bsB\bsvarphi,
    \;\forall \bsphi
}
With this we obtain that any $f$ verifying~\ref{eq:fphi} is such that  
$$f(\bsphi)=f(\Cdag\Clot\bsphi-\bsB\bsvarphi)= f(\Cdag\Clot\bsphi)=
g(\Clot\bsphi).$$

\section{Wrapped kernel of operator $\Clot$}\label{app:intamb}
The kernel of operator $\Clot$ is given by $\ker\Clot=\IMA\bsB$
(equation~\ref{lannes}). With dimensional arguments, it is easy to see that 
\beas{\IMA\bsB=\IMA\bsBb}
where $\bsBb$ is obtained by removing the first column of operator $\bsB$, so
we have\beq{\ker\Clot=\IMA\bsBb\label{eq:kerc}}

Let us now characterize the set of $\eker{\bsphi}$ such that :
$$\Clot\eker{\bsphi}\equiv 0 \,[2\pi]$$
Because $\Clot$ has integer components, $\eker{\bsphi}$ can be considered
modulo $2\pi$.
With equation~\ref{eq:kerc}, we obtain:
\beq{\label{eqb}\exists \bsalpha_1,\,\eker{\bsphi}\equiv \Cdag\left(0
  \,[2\pi]\right)+\bsBb\bsalpha_1\,[2\pi]}
Because $\bsBb$ has integer components, $\bsalpha_1$ can be considered
modulo $2\pi$. The issue here is to evaluate the $\Cdag\left(0
  \,[2\pi]\right)$ term, i.e. the value of $\Cdag\left(2\pi\bskappa\right)$,
with $\bskappa$ any integer vector.\\

Equations~\ref{props} show that $\Clot=\left[\begin{array}{c|c}\bsM&\Id\end{array}\right]$.
The integer vector
$\bsmu\defeq\left[\begin{array}{c}\bs0\\\bskappa\end{array}\right]$ is then such
that $$\Clot\bsmu=\left[\begin{array}{c|c}\ast&\Id\end{array}\right]\left[\begin{array}{c}\bs0\\\bskappa\end{array}\right]=\bskappa.$$
Then we have:
\begin{align*}
\Clot\bsmu=\bskappa
\so & \Clot\bsmu'=\Clot\Clot^\dag\bskappa\\
\so & \Clot(\Clot^\dag\bskappa-\bsmu)=0\\
\so & \exists\bsalpha_2,\;\Clot^\dag\bskappa-\bsmu=\bsB\bsalpha_2\\
\so & \exists\bsalpha_2,\;2\pi\Clot^\dag\bskappa=2\pi\bsmu+\bsB(2\pi\bsalpha_2)\\
\so & \exists\bsalpha_2,\;\Cdag\left(0\,[2\pi]\right)+\bsBb\bsalpha_1 \equiv \bsB(\underbrace{2\pi\bsalpha_2+\bsalpha_1}_{\bsalpha})\,[2\pi].
\end{align*}

So equation~\ref{eqb} yields
\beq{\label{eqc}\exists \bsalpha,\,\eker{\bsphi}\equiv \bsBb\bsalpha\,[2\pi]}

\section{Square-root of a Gaussian distribution\label{app:racgauss}}
Let us suppose we measure the squared value $s$ of a positive value $a$, with
an additive Gaussian noise:
\beq{\label{eqds}\data{s}=a^2+\noise{s},}
$\noise{s}$ being 0 mean Gaussian with the variance $\sigma_s^2$. Let $\hat{a}$
be the estimator of $a$ from $\data{s}$ defined by
$$\hat{a}=\left\{\begin{array}{c}\sqrt{\data{s}}, \;\RM{ if }\;\data{s}>0\\
    0\;\RM{else}\end{array}\right.$$ 
$\hat{a}$ can be seen as pseudo-data. The data model of $\hat{a}$ derived from
equation~\ref{eqds} is not
additive Gaussian. As will be shown in section \ref{app:KL}, an optimal
Gaussian approximation of the data model of $\hat{a}$ would be:
\beq{\label{eqda}\hat{a}=a+\noise{a},}
with $\noise{a}$ a Gaussian noise with a mean equal to $<\hat{a}>$ and a
standard deviation $\sqrt{\var(\hat{a})}$. 

\begin{figure}[h!]
\begin{center}
\includegraphics[width=.7\linewidth]{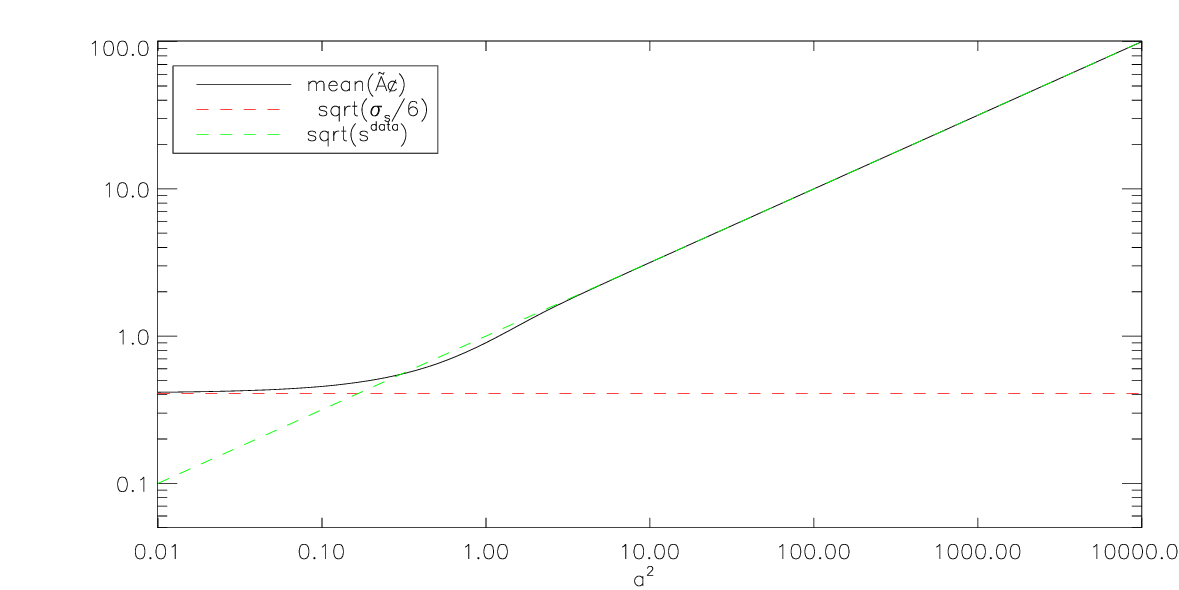}
\caption{Behavior of $<\hat{a}>$ in function of  $a^2$ with a unit $\sigma_s$\label{fig:racgauss2}}
\end{center}
\end{figure}

\begin{figure}[h!]
\begin{center}
\includegraphics[width=.7\linewidth]{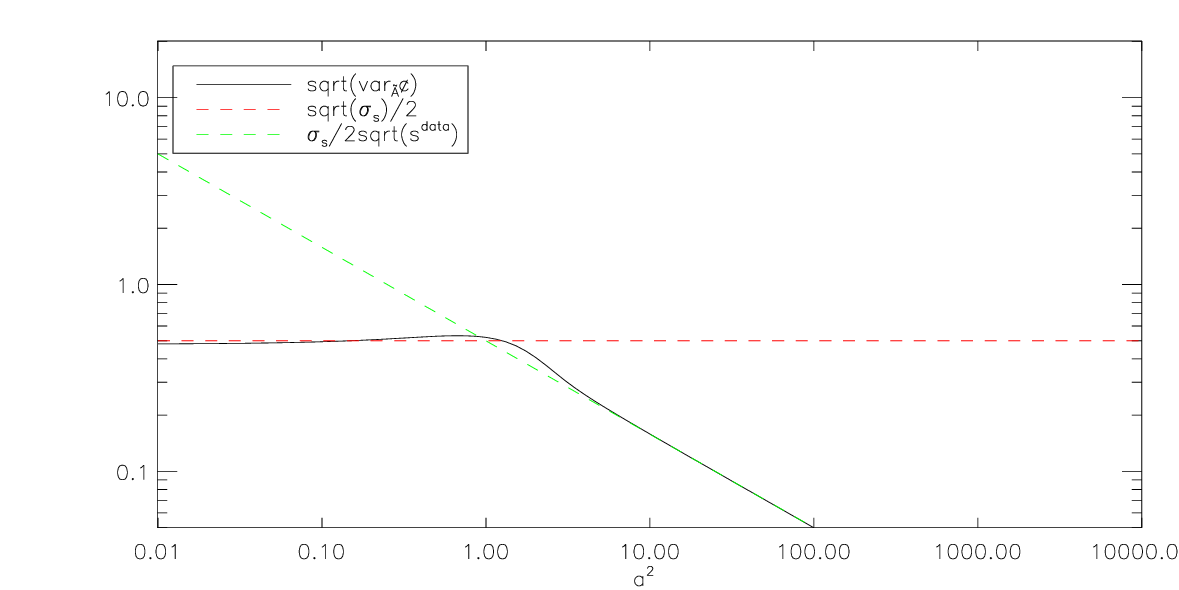}
\caption{Behavior of $\sqrt{\var(\hat{a})}$ in function of  $a^2$ with a unit $\sigma_s$ \label{fig:racgauss3}}
\end{center}
\end{figure}

We have studied the behavior of the mean $<\hat{a}>$ and standard deviation
$\sqrt{\var(\hat{a})}$ of this estimator for various values of $a^2$, with a unit $\sigma_s$ (see
figs.~\ref{fig:racgauss2} and~\ref{fig:racgauss3}). 

We can distinguish two regimes for $<\hat{a}>$:
\begit{
\item a low mean regime, where $a^2\le\sigma_s/6$ : a non negligible part of
  the distribution of $\data{s}$ around $a^2$ is in the negative domain.
  Because $\hat{a}$ estimates a null value for $a$ when $\data{s}$ is
  negative, its mean will mainly depend on the width of the Gaussian wings. A
  good approximation of $<\hat{a}>$ is $\sqrt{\sigma_s/6}$;
\item a high mean regime, where $a^2\ge\sigma_s/6$ : the most part of
  the distribution of $\data{s}$ around $a^2$ is in the positive domain.
  The fact that $\hat{a}$ estimates a null value for $a$ when $\data{s}<0$
  does not impact its mean $<\hat{a}>$, which is close to $a$. Because $a$ is
  not known, we choose $<\hat{a}>=\sqrt{\data{s}}$.
}

We can distinguish the same two regimes for $\sqrt{\var(\hat{a})}$. However, the transition is around $\sigma_s$:
\begit{
\item when $a^2\le\sigma_s$, the fact that $\hat{a}$ estimates a null value for $a$ when $\data{s}$ is
  negative tends to diminish its standard deviation, which we approximate by
$\sqrt{\var(\hat{a})}\simeq \sqrt{\sigma_s}/2$;
\item in the high mean regime, where $a^2\ge\sigma_s$, the most part of
  the distribution of $\data{s}$ around $a^2$ is in the positive domain, and
  $\sqrt{\var(\hat{a})}$ is close to the classical expression. This expression corresponds
  to a first order expansion in $\sigma_a$:
$$(a+\sigma_a)^2=a^2+\sigma_s \so 2a\sigma_a\simeq\sigma_s.$$} $\sigma_s/2a$.
Because $a$ is
  not known, we choose $\sqrt{\var(\hat{a})}=\sigma_s/2\sqrt{\data{s}}$.

We then propose the pseudo-data model $$\data{a}=a+\noise{a}$$

with $\data{a}=\left\{\begin{array}{c}\sqrt{\data{s}}, \;\RM{ if }\;\data{s}>0\\
    0\;\RM{else}\end{array}\right.$ and $ \noise{a}$ a Gaussian noise with
mean and standard deviation defined by
with :
\beas{
\bar{a}&=\left\{\begin{array}{rl}
\sqrt{\sigma_s/6}, \;&\RM{ if }\;\data{s}\le\sigma_s/6\\
\sqrt{\data{s}}, \;&\RM{ if }\;\data{s}\ge\sigma_s/6\end{array}\right.\\
\sigma_a&=\left\{\begin{array}{rl}\sqrt{\sigma_s}/2 \;&\RM{ if }\;\data{s}\le \sigma_s\\
\frac{\sigma_s}{2\sqrt{\data{s}}}, \;&\RM{ if }\;\data{s}\ge\sigma_s\end{array}\right.
}
 We also decide to discard the data such that $\data{s}\le -\sigma_s $.

\section{Cartesian Gaussian approximation to a polar Gaussian
  distribution\label{app:KL}}

If we define \bea{\yba(\bsx,t)\defeq {\bsH(t)}\bsx\cdot\EXP^{\I\bsBb\bsalpha(t)},} equation~(\ref{modpol}) reads
\bea{\label{modpol2}\left\{\bead{
      \data{\bsa}(t)&=\abs{\yba}(\bsx,t)+\bruit{\bsa}(t),\;\;\;&\bruit{\bsa}(t)\gauss{\zerob}{\bsR_{\bsa(t)}}.\\
      \data{\bsphi}(t)&\congr
      \arg{\yba}(\bsx,t)+\bruit{\bsphi}(t),\;\;\;&\bruit{\bsphi}(t)\gauss{\bs0}{\bsR_{\bsphi(t)}}.}\right.}

\subsection{General expression}
With consider a polar distribution of a Gaussian vector $\bsy$ of modulus
$\bsa$ and phase $\bsphi$:
\bea{\label{model_myope}
\data{\bsphi}&=\bar{\bsphi}+\bruit{\bsphi}\\
\data{\bsa}&=\bar{\bsa}+\bruit{\bsa}}
where $\bruit{\bsphi}$ and $\bruit{\bsa}$ are 0 mean real Gaussian vectors, of
covariance matrices $\cov{\bsa}$ and $\cov{\bsphi}$ (the vectors
$\bruit{\bsphi}$ and $\bruit{\bsa}$ are supposed uncorrelated).

With the definitions
\bea{\label{eq:defs}\left\{
\begin{aligned}
\bar{\bsy}&\defeq\bar{\bsa}\exp\I\bar{\bsphi}\\
\bruit{\bsy}&\defeq\data{\bsy}-\bar{\bsy}\\
\brad{\bsy}&\defeq\RE{\bruit{\bsy}\EXP^{-\I\bar{\bsphi}}}\\
\btan{\bsy}&\defeq\IM{\bruit{\bsy}\EXP^{-\I\bar{\bsphi}}}\\
\bbys&\defeq\bby
\end{aligned}\right.}
we gather:
\bea{\label{eq:defsRT}\left\{
\begin{aligned}
\brad{\bsy}&=\left[\bar{\bsa}+\bruit{\bsa}\right]\cos\bruit{\bsphi}
-\bar{\bsa}\\
\btan{\bsy}&=\left[\bar{\bsa}+\bruit{\bsa}\right]\sin\bruit{\bsphi}
\end{aligned}\right.}
A complex vector is Gaussian if and only if each of its components is Gaussian. A complex
is Gaussian if and only if, in any Cartesian basis, its two components are
Gaussian. So $\bsy$ is Gaussian if and only if $\bbys$ is Gaussian, which is
not the case\citeb{Meimon-a-05}. In what follows, we show how to optimally approximate the
distribution of $\bbys$ by a Gaussian distribution.

\subsection{Gaussian Approximation }
We characterize our Cartesian additive Gaussian approximation, \ie its mean
$\Avg{\bbys}$ and
covariance $\cov{\bbys}$, by minimizing the Kullback-Leibler distance between the two
noise distributions, which gives \citeb{Meimon-a-05}:
\bea{\label{eq:KL}\left\{
\begin{aligned}
\Avg{\bbys}&=\Esp{\bby}=\left[\begin{array}{c}\brad{\bar{\bsy}}\\\btan{\bar{\bsy}}\end{array}\right]\\
\cov{\bbys}&=\Esp{\left[\begin{array}{c}\brad{\bar{\bsy}}-\brad{{\bsy}}\\\btan{\bar{\bsy}}-\btan{{\bsy}}\end{array}\right]
\left[\begin{array}{c}\brad{\bar{\bsy}}-\brad{{\bsy}}\\\btan{\bar{\bsy}}-\btan{{\bsy}}\end{array}\right]\T}
\end{aligned}\right.}
and we define \beas{
  \cov{\bbys}\defeq\left[\begin{array}{cc}\cov{\RM{rad,rad}}&\cov{\RM{rad,tan}}\\\cov{\RM{rad,tan}}\T&\cov{\RM{tan,tan}}\end{array}\right]
}

For a 0 mean Gaussian vector $\bruit{\bsphi}$ of
covariance matrix $\cov{\bsphi}$,

\bea{\label{eq:espsin}\begin{aligned}
\Esp{\sin \bruit{\phi}_i}&=0\\
\Esp{\cos \bruit{\phi}_i}&=\exp-\frac{{\cov{\bsphi}}_{ii}}{2}\\
\Esp{\sin\bruit{\phi}_i\sin\bruit{\phi}_j}&=\sinh{\cov{\bsphi}}_{ij}\cdot\exp-\frac{{\cov{\bsphi}}_{ii}+{\cov{\bsphi}}_{jj}}{2}\\
\Esp{\cos\bruit{\phi}_i\cos\bruit{\phi}_j}&=\cosh{\cov{\bsphi}}_{ij}\cdot\exp-\frac{{\cov{\bsphi}}_{ii}+{\cov{\bsphi}}_{jj}}{2}\\
\Esp{\cos\bruit{\phi}_i\sin\bruit{\phi}_j}&=0
\end{aligned}}

By combining equations.~\ref{eq:KL},~\ref{eq:defs},~\ref{eq:defsRT} and~\ref{eq:espsin}, we obtain:


\bea{\label{eq:covgauss}\begin{aligned}
\Esp{\brad{\bsy}_i}&=\bar{a}_i\left[\EXP^{-\frac{{\cov{\bsphi}}_{ii}}{2}}-1\right]\\
\Esp{\btan{\bsy}_i}&=0\\
\left[\cov{\RM{rad,rad}}\right]_{ij}&=\left[\bar{a}_i\bar{a}_j\left(\cosh{\cov{\bsphi}}_{ij}-1\right)+\cov{a_{ij}}\cosh{\cov{\bsphi}}_{ij}\right]
\cdot\EXP^{-\frac{{\cov{\bsphi}}_{ii}+{\cov{\bsphi}}_{jj}}{2}}\\
\left[\cov{\RM{rad,tan}}\right]_{ij}&=0\\
\left[\cov{\RM{tan,tan}}\right]_{ij}&=\left(\bar{a}_i\bar{a}_j+\cov{a_{ij}}\right)\sinh{\cov{\bsphi}}_{ij}
\cdot\EXP^{-\frac{{\cov{\bsphi}}_{ii}+{\cov{\bsphi}}_{jj}}{2}}
\end{aligned}}
\subsection{The scalar case}
Now, we make the additional assumption that both $\bruit{\bsphi}$ and
$\bruit{\bsa}$ are decorrelated, i.e. 
\beas{\left\{
\begin{aligned}
\cov{\bsa}&=\diag{\sigma_{a,i}^2}\\
\cov{\bsphi}&=\diag{\sigma_{\phi,i}^2}
\end{aligned}\right.}

We obtain:
\beas{\left\{
\begin{aligned}
\cov{\RM{rad,rad}}&=\diag{\sigma_{rad,i}^2}\\
\cov{\RM{tan,tan}}&=\diag{\sigma_{tan,i}^2}\\
\cov{\RM{rad,tan}}&=0
\end{aligned}\right.}
with
\bea{\label{eq:covgauss2}\begin{aligned}
\sigma_{rad,i}^2&=\frac{\bar{a}^2_i}{2}\left(1-\EXP^{-\sigma_{\phi,i}^2}\right)^2+
\frac{\sigma_{a,i}^2}{2}\left(1+\EXP^{-2\sigma_{\phi,i}^2}\right)\\
\sigma_{tan,i}^2&=\frac{\bar{a}^2_i}{2}\left(1-\EXP^{-2\sigma_{\phi,i}^2}\right)+
\frac{\sigma_{a,i}^2}{2}\left(1-\EXP^{-2\sigma_{\phi,i}^2}\right)
\end{aligned}}

In this case, we can plot for one complex visibility the true noise
distribution - i.e. a Gaussian noise in phase and modulus - and our Gaussian
approximation (see fig.~\ref{fig:bananneandell}).

\section{Acknowledgments}

The authors want to express their special thanks to Eric Thi\'ebaut for his
support and for letting them use his minimization software. Serge Meimon is
very gratefull to Guy Perrin and Sylvestre Lacour, who allowed him to
participate to two IOTA observing campaigns. We also would like to thank all
the people who contributed to the existence and success of the IOTA
interferometer, in particular John Monnier, Wes Traub, Jean-Philippe Berger
and Marc Lacasse. Serge Meimon also thanks Vincent Bix Josso for his help on
appendix B. Serge Meimon and Laurent Mugnier acknowledge support from PHASE,
the space and ground based high angular resolution partnership between ONERA,
Observatoire de Paris, CNRS and University Denis Diderot Paris 7.

\begin{sloppypar}
\enlargethispage{\baselineskip}
Corresponding author Serge Meimon can be reached at
\texttt{Serge.Meimon@onera.fr}
\end{sloppypar}

\providecommand{\inpreparationname}{en pr\'eparation}
  \providecommand{\submittedname}{soumis}
  \providecommand{\acceptedname}{accept\'e pour publication}
  \providecommand{\tobepublishedname}{\`a para\^{\i}tre}
  \providecommand{\contractname}{Contrat}
  \providecommand{\conferencedatename}{Date conf\'erence~: }
  \providecommand{\patent}[2]{Brevet #1 #2}
  \providecommand{\firstabbrevname}{1\textsuperscript{\`ere} }
  \providecommand{\secondabbrevname}{2\textsuperscript{\`eme} }
  \providecommand{\thirdabbrevname}{3\textsuperscript{\`eme} }
  \providecommand{\fourthabbrevname}{4\textsuperscript{\`eme} }
  \providecommand{\fifththabbrevname}{5\textsuperscript{\`eme} }
  \providecommand{\sixththabbrevname}{6\textsuperscript{\`eme} }
  \providecommand{\inpreparationname}{en pr\'eparation}
  \providecommand{\submittedname}{soumis}
  \providecommand{\acceptedname}{accept\'e pour publication}
  \providecommand{\tobepublishedname}{\`a para\^{\i}tre}
  \providecommand{\contractname}{Contrat}



\newpage

\listoffigures
\end{document}